# Spin Polarization Phenomena and Pseudospin Quantum Hall Ferromagnetism in the HgTe Quantum Well

M. V. Yakunin[a], A. V. Suslov[b], S. M. Podgornykh[a], S. A. Dvoretsky[c] and N. N. Mikhailov[c]

[a]*Institute of Metal Physics, S. Kovalevskaya Str., 18, Ekaterinburg 620990, Russia*
[b]*NHMFL, FSU, 1800 East Paul Dirac Drive, Tallahassee, Florida 32310, USA*
[c]*Institute of Semiconductor Physics, Lavrentyev Ave., 13, Novosibirsk 630090, Russia*

**Abstract.** The parallel field of a full spin polarization of the electron gas in a $\Gamma_8$ conduction band of the HgTe quantum well was obtained from the magnetoresistance by three different ways in a zero and quasi-classical range of perpendicular field component $B_\perp$. In the quantum Hall range of $B_\perp$ the spin polarization manifests in anticrossings of magnetic levels, which were found to strongly nonmonotonously depend on $B_\perp$.

Spin polarization underlies the basic principles of spintronics devices. Full spin polarization of the electronic system is easily achievable in the HgTe conduction band due to its large value of $g^*m^*/m_0$, the effective Lande $g$-factor multiplied by the effective to free mass ratio, thus making it promising for applications and studies of a variety of spin phenomena. In particular, rich patterns of the spin level coincidences are observed in the HgTe quantum well (QW) under tilted magnetic fields that extend into the range of high field component $B_\perp$ perpendicular to the layer where the quantum Hall effect (QHE) is well realized [1].

We present a study of quantum magnetotransport under tilted magnetic fields in a 20.3 nm wide HgTe symmetric QW with the electron gas density $n_S \approx 1.5\times10^{15}$ m$^{-2}$ and mobility of 22 m$^2$/V·s. The spin level coincidences manifest in a number of peculiarities in the longitudinal $\rho_{xx}(B_\perp,B_\parallel)$ (Fig. 1a, $B_\parallel$ – field component parallel to the layer) and Hall magnetoresistivities (MR) residing on different trajectories in the $(B_\perp,B_\parallel)$-plane. First, they align well on a set of straight beams going from zero at fixed tilt angles $\theta_r$ that satisfy a typical relation for the coincidences: $g^*m^*/m_0 = 2r\cdot\cos\theta_r$, $r = 1,2,3…$[2]. This means that the coincidences in the $\Gamma_8$ conduction band in our case may be well described in terms of a usual $\Gamma_6$ band as in a traditional semiconductor QW. Second, another system of traces along the coincidences, descending from a single point $B_\parallel^0 = 15.4$ T on the $B_\parallel$ axis, may be drawn. This series is approximately described at high enough tilts in terms of the $\Gamma_6$-like band by the equation $B_\parallel = 2(B_1 - MB_\perp)/(g^*m^*/m_0)$, $M = 1,3,5…$, where $B_1 = hn_S/e$ is the MR minimum position for the magnetic level filling factor $v = 1$, in a good agreement with our experimental data. The convergence point $B_\parallel^0 = 2B_1/(g^*m^*/m_0)$ fulfil the relation $g^*\mu_B B_\parallel^0 = 2E_F$, with $E_F$ – the Fermi level, $\mu_B$ – Bohr magneton, thus yielding the field of a full spin polarization of the electron gas under pure parallel field (Fig. 1b).

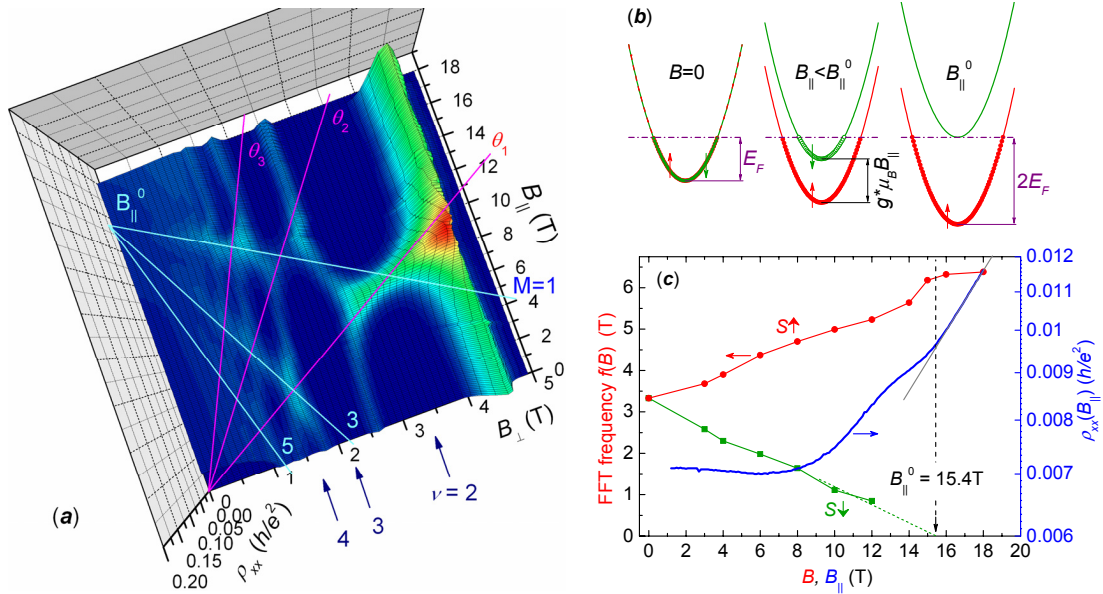

**FIGURE 1.** (a) The magnetoresistivity $\rho_{xx}(B_\perp, B_\parallel)$ at 0.32 K after intermediate illumination. (b) The evolution of spin splitting under pure parallel field until all the electrons pass into the lower spin subband acquiring the same spin polarization. (c) Three ways of probing the full spin polarization by means of magnetoresistivity: the convergence point $B_\parallel^0$ of the descending trajectories on the $(B_\perp, B_\parallel)$-plane (the dashed vertical); the FFT frequency as a function of $B$: falling of the lower ($S\downarrow$) line to zero and a concomitant saturation of the upper ($S\uparrow$) line; a point on $\rho(B = B_\parallel)$ separating two functions.

The other way to estimate the redistribution of electrons between two spin subbands is from the Fourier analysis of oscillations in $\rho_{xx}(1/B_\perp)$ taken along the circle trajectories in the $(B_\perp, B_\parallel)$-plane for fixed values of the total field $B$. This results in two lines of FFT frequency $f$ vs. $B$ (Fig. 1c) describing the behavior of electron densities in the subbands: $n_{si} = f_i \times e/h$. The lines diverge from a single point with the lower line going to zero (exhausting of the upper subband) and the upper line simultaneously going to saturation. This result is compared with MR in a pure $B_\parallel$ (Fig. 1c). It appears that just this field separates two kinds of dependences in $\rho(B = B_\parallel)$: a somewhat complicated function at lower fields from a monotonously increasing one at higher fields as has been seen in a number of studies conducted in parallel fields [3]. Thus, all three techniques indicate a full spin polarization occurring at the same field.

At higher $B_\perp$, where the QHE is well developed, the MR features for coincidences acquire a complicated structure with local $\rho_{xx}$ peaks splitted in couples of peaks shifted in opposite directions of $B_\perp$ (Fig. 1a), thus indicating the formation of anticrossings at the points of expected level crossings. The effect is significantly enhanced after IR illumination of the sample that causes a considerable improvement of oscillations indicating a narrowing of magnetic levels. Unexpectedly, it was found that the anticrossings depend nonmonotonously on $B_\perp$: the anticrossing at $v = 3$ manifests pronouncedly stronger than the neighboring ones at $v = 2$ and 4, as it is easily seen while going along the descending trace for $M = 1$ on the $(B_\perp, B_\parallel)$-plane. The difference is dramatically enhanced after illumination and with decreased temperature. The activation gaps deduced from the temperature dependences of MR at anticrossings confirm this nonmonotonicity with the $\nu = 3$ gap being a half an order larger than those for its neighbors. This result looks counterintuitive since the overlapping of magnetic levels with decreased $B_\perp$ seems to monotonously destroy the reasons for appearing of anticrossings as it has been observed so far on other materials [4,5].

The conventional explanation of the anticrossings is in terms of the pseudospin anisotropy of the electronic system [6]: as the approaching magnetic levels (having different pseudospin numbers) tend to swop their order in energy relatively $E_F$, the Hartree-Fock energy of the system may decrease and, as this decrease starts before the level crossing (due to a hybridization of the levels), this crossing does not occur, the stronger is the energy gain the larger is the anticrossing gap. The systems inclined to transitions into pseudospin ordered states under QHE conditions are called QH Ferromagnets (QHF). Estimations for an easy-axis QHF in a 2D layer according to Eq. 21 in [6] does not yield a big difference for anticrossings at $\nu = 2, 3$, and 4. Therefore, we tentatively attribute the observed difference to the coupling of $B_\parallel$ with the orbital degree of freedom in a QW of a finite width resulted in a substantial difference in the charge density profiles across the QW for the two pseudospin levels [5]. This coupling enhances the magnetic anisotropy energy for $\nu = 3$ coincidence as compared to that for $\nu = 2$ since its $B_\parallel$ is about a factor of 1.5 stronger, causing a relative shrinkage of the wave functions. On the other hand, the coincidences at $\nu \geq 4$ are restored because they go outside of the QH range of $B_\perp$. The observed sharp changes of anticrossings with the fields and, for fixed anticrossings, with a change in a sample state due to illumination indicate the phase-transition-like character of these transformations.

## ACKNOWLEDGMENTS

Authors are grateful to E. Palm, T. Murphy, J. H. Park, and G. Jones for help with the experiment. Supported by RFBR, projects 11-02-00427, 09-02-96518. NHMFL is supported by NSF (DMR-0654118), the State of Florida, and the US DOE.

## REFERENCES

1. M. V. Yakunin, S. M. Podgornykh, N. N. Mikhailov and S. A. Dvoretsky, *Physica E* **42**, 948-951 (2010); M. V. Yakunin *et al.*, *J. Phys.: Conf. Ser.* (2011), *in print*.
2. R. J. Nicholas, M. A. Brummell, J. C. Portal *et al.*, *Solid State Comm.* **45,** 911-914 (1983); R .J. Nicholas, R. J. Haug, K. von Klitzing and G. Weinmann, *Phys. Rev. B* **37**, 1294-1302 (1988).
3. J. Yoon, C. C. Li, D. Shahar, D. C. Tsui and M. Shayegan, *Phys. Rev. Lett.* **84**, 4421-4424 (2000); B. Nedniyom, R. J. Nicholas, M. T. Emeny, L. Buckle, A. M. Gilbertson, P. D. Buckle and T. Ashley, *Phys. Rev. B* **80**, 125328 (2009); I. L. Drichko, I. Yu. Smirnov, A. V. Suslov, O. A. Mironov and D. R. Leadley., *Phys. Rev. B* **79**, 205310 (2009).
4. S. Koch, R. J. Haug, K. V. Klitzing and M. Razeghi, *Phys. Rev. B* **47**, 4048-4051 (1993); W. Desrat, F. Giazotto, V. Pellegrini, M. Governale, F. Beltram *et al.*, *Phys. Rev. B* **71**, 153314 (1993).
5. T. Jungwirth, S. P. Shukla, L. Smrcka, M. Shayegan and A. H. MacDonald, *Phys. Rev. Lett.* **81**, 2328-2331 (1998).
6. T. Jungwirth and A. H. MacDonald, *Phys. Rev. B* **63**, 035305 (2000).